\documentstyle[twoside,fleqn,psfig,espcrc2]{article}

\newcommand{\be}{\begin{equation}}
\newcommand{\ee}{\end{equation}}
\newcommand{\bea}{\begin{eqnarray}}
\newcommand{\eea}{\end{eqnarray}}
\newcommand{\no}{\noindent}
\newcommand{\e}{{\rm e}}

\newlength{\figwidth}
\newlength{\figheight}
\title{Finite Temperature QCD on Anisotropic Lattices
 \thanks{Talks by H. Matsufuru and I.-O. Stamatescu at LATTICE98. We thank
JSPS, DFG and the European Network ``Finite Temperature Phase Transitions in Particle Physics" for support.}}

\author{QCD-TARO:
Ph.~de~Forcrand%
\address{SCSC, ETH-Z\"urich, CH-8092 Z\"urich, Switzerland \\
$^{\rm b}$Dept. F\'{\i}sica Te\'orica, Universidad Aut\'onoma de Madrid,
      E-28049 Madrid, Spain \\
$^{\rm c}$Dept. of Appl. Phys., Fac. of Engineering,
           Fukui Univ., Fukui 910-8507, Japan \\
$^{\rm d}$Dept. of Physics, Tezukayama Univ.,Nara 631-8501, Japan \\
$^{\rm e}$Institut f\"ur Theoretische Physik, Univ. Heidelberg
           D-69120 Heidelberg, Germany \\
$^{\rm f}$Dept. of Physics, Hiroshima Univ.,
           Higashi-Hiroshima 739-8526, Japan \\
$^{\rm g}$Res. Inst. for Inform. Sc. and Education, Hiroshima Univ.,
           Higashi-Hiroshima  739-8521, Japan \\
$^{\rm h}$FEST, Schmeilweg 5, D-69118 Heidelberg, Germany \\
$^{\rm i}$Hiroshima University of Economics, Hiroshima 731-01, Japan },
M.~Garc{\'\i}a~P\'erez$^{\rm b}$,
T.~Hashimoto$^{\rm c}$,
S.~Hioki$^{\rm d}$,
H.~Matsufuru$^{\rm ef}$,
O.~Miyamura$^{\rm f}$,
A.~Nakamura$^{\rm g}$,
I.-O.~Stamatescu$^{\rm eh}$,
T. Takaishi$^{\rm i}$,
and
T.~Umeda$^{\rm f}$
}

\begin{document}
\begin{abstract} We present results for mesonic propagators in temporal and
spatial direction and for topological properties at $T $ below and above the deconfining 
transition in quenched QCD.  We use anisotropic lattices 
and Wilson fermions.
\end{abstract}
\maketitle

\section{Questions concerning QCD at $T > 0$}

With increasing temperature we expect the physical picture promoted by QCD 
to change according to a phase transition. Questions related with this
transition are:  whether chiral symmetry restoration 
and deconfinement 
represent the same phenomenon or succeed by very near but different 
transitions; how can one characterize the change in the vacuum structure 
(topology, monopoles, vortices, etc); which are the precise changes in the
properties of hadrons; etc. 
The present paper reports the results from a project going on over a number 
of years and analyzing some of these questions. Although this 
analysis is not closed, the present stage achieves a certain 
description which we shall present here
together with the 
physical picture suggested by it and further problems.

The primary question  in this analysis has been the
structure of the hadron correlators and its dependence on temperature.
A second question raised in the later phase of the project 
concerns the relation to the topological properties. Both these problems
have been treated in quenched QCD. Both these questions will be 
pursued to completion. We also intend to extend preliminary studies 
for full QCD to a systematic analysis.

Since $T>0$ breaks Lorentz symmetry the euclidean formulation of the 
finite temperature problem specifies a ``temperature" (euclidean time) 
axis (see, e.g, \cite{tgen}). Therefore we need to investigate hadronic
correlators with full ``space-time" structure, in particular the
propagation in the euclidean time direction.  The latter, however, puts
special problems because of the inherently limited physical length of
the lattice,
${\it l}_{\tau} = {1 \over T}$. To preserve at least a reasonably fine 
discretization of the time axis\footnote{In the following we shall
always refer to euclidean time, denoted $t$, unless explicitly stated otherwise.}
(needed in order to follow the details of
the time-dependence of the correlators) 
we chose to use anisotropic lattices with $a_{\sigma}/a_{\tau}=\xi >1 $
which can ensure the above requirement without increasing  
too much the total volume of the lattice (section 2). 
To separate ground and excited states 
we need a more refined definition of the hadronic propagators.
 The strategy used will be described in 
section 3. Sections 4 and 5 present our main results for the hadronic 
correlators and section 6 discusses the temperature dependence of the
topological properties of the configurations. Section 7 is reserved for 
conclusions and outlook.

\section{$T>0$ QCD on anisotropic lattices}

Anisotropic lattices are realized by introducing different space-space
and space-time couplings, e.g. for the Wilson Yang-Mills action:
\be
S_{YM}=-{{\beta} \over 3}\left({1\over\gamma }Re\hbox{Tr}
\Box_{\sigma\sigma}+\gamma Re\hbox{Tr}
\Box_{\sigma \tau}
\right)
\label{e.act}
\ee
\no This produces a cut-off anisotropy $\xi = \gamma \eta (\beta,\gamma)$, the temperature being given by
  $ T = {1 \over {{\it l}_{\tau}}} = {{\xi} \over {N_{\tau} a_{\sigma}}}$.

The function $ \eta (\beta,\gamma)$ depends on the dynamics. To 
determine it correlation functions in different directions  are calculated for given $\beta, \gamma$
at $T=0$ and required to show
isotropy under rescaling of the time distances by $\xi$ (``calibration"):
\begin{equation}
F_n^{\sigma}(z) = F_n^{\tau}(t = \xi z)
\label{e.cal}
\end{equation}
\no Once the Yang-Mills calibration has been performed, we calculate
hadron correlators from the action
 ($T_{\mu}$ are lattice translation operators):
\bea
S_F = 2 \kappa_{\sigma} {\bar{\Psi}} W \Psi,\  W = 1 - \\ 
\kappa_{\sigma}\left(\sum_i \Gamma_i^+ U_i T_i +
\gamma_F\Gamma_4^+ U_4 T_4 \right) 
+\hbox{``h.c."},\\
\kappa_{\sigma}^{-1} =
     2(m_0 +3 + \gamma_F) = \kappa^{-1}+\gamma_F -1, \label{e.kappa} \\
\Gamma_{\mu}^{\pm} =
1 \pm \gamma_{\mu},\ \ \gamma_{\mu}^2 = 1 \nonumber
\eea
\no and tune $\gamma_F$ to obtain the same $\xi$ when the physical
isotropy condition  eq. (\ref{e.cal}) is applied to these correlators.
If the action leads to strong artifacts, (\ref{e.cal}) cannot 
be fulfilled simultaneously for all observables and 
the result depends in particular on the action used. In \cite{biel1}
we compared the calibration for various actions, see also \cite{Sakai98}. 

Our quenched QCD analysis uses Wilson action with $\beta = 5.68$ and $\gamma=4$
on lattices of $12^3 \times N_{\tau}$  with $N_{\tau}\  =\  72,\  20,\  16$ and
$12$. We have  $a_{\sigma}\simeq 0.24~$fm. The critical temperature is found from the behavior of the 
Polyakov loop susceptibility to be at $N_{\tau}$ slightly above 18, which gives
for the  above lattices the temperatures $T \ \simeq \ 0,\ 0.93 T_c,\ 1.15 T_c$ and 
$1.5T_c$ (see \cite{TARO96}).  
The results to be reported below correspond to two sets of analysis:\par
\no - {\it Set-A} uses Wilson loops to perform the Yang-Mills calibration. We 
find a value for $\xi$ ranging between 
5.3 for wide Wilson loops in eq. (\ref{e.cal}) and 6.3 for thin Wilson loops. For the analysis an average $\xi = 5.9$ is used and we find $\gamma_F = 5.4$.  Because of the ambiguity in 
$\xi$ we use this set only to investigate 
the temperature dependence of the time-correlators but did not attempt to 
make a systematic comparison with space (or ``screening") correlators.\par
\no - {\it Set-B} corresponds to a calibration performed for the static 
quark potential, which is a more adequate physical observable
for requiring isotropy than the Wilson loops themselves \cite{biel1}. 
The ambiguity in fixing $\xi$ is reduced, we 
find a $\xi\simeq 5.3$ and we can compare correlators corresponding to 
both time and space propagation. \par
\begin{figure}[tb]
\vspace*{-0.28cm}
\leavevmode\psfig{file=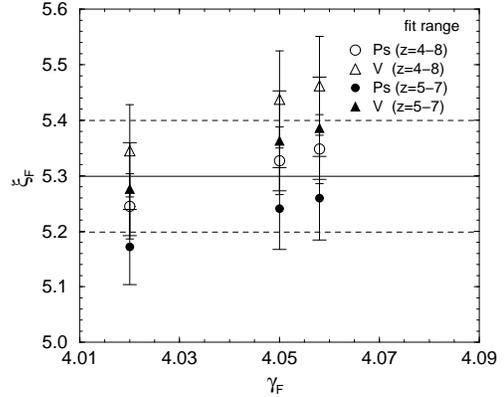,width=\figwidth}
\vspace{-2.4cm} \\
\caption{
Fermionic calibration.
Three values of $\gamma_F$ are tried with fixed
$\kappa_{\sigma}=0.081$, and the value $\gamma_F=4.05$
is adopted. }
\label{f.cal}
\vspace{-0.6cm}
\end{figure}

\no In Fig. \ref{f.cal} we present
the calibration for {\it Set-B}.
The various parameters are given in
the table.

\begin{table*}[t]
\begin{center}
\begin{tabular}{cccccccccc}
\hline
set & $\xi$ & $a_{\sigma}^{-1}$ GeV& $\kappa_{\sigma}$  &
$\gamma_F$ & \#conf & $m_{PS}$ & $m_{V}$ & $a$ (eq.(9))& $p$ (eq.(9))\\ 
\hline
A & 5.9 & $\sim$ 0.81 & 0.068 & 5.4  & 20 & 0.1071(27) & 0.1311(31) &
  0.4346(33) & 1.2635(93) \\
\hline
B & 5.3(1)&  0.85(3)  & 0.081 & 4.05 & 30 & 0.1799(18) & 0.1992(22) &
  0.3785(33) & 1.2889(78) \\
  &       &           & 0.084 & 3.89 & 30 & 0.1510(19) & 0.1772(24) &
  0.3797(31) & 1.2767(75) \\
  &       &           & 0.086 & 3.78 & 30 & 0.1355(19) & 0.1664(26) &
  0.3800(25) & 1.2634(75) \\
\hline
\end{tabular}
\end{center}
%\caption{The parameters of the simulations. }
\label{tab:params}
\vspace{-0.5cm}
\end{table*}

\section{Strategy}

Increasing the temperature is expected to induce significant changes in the 
structure of the hadrons. In particular, the mesonic peak in
the spectral function of the correlators 
may move and develop a width. The question
is, how to observe this change by distinguishing it from the 
influence of higher excitations which will not be dumped enough at
the time scale of the inverse temperature.  

Two pictures are frequently used to describe the 
intermediate and the high temperature regimes: the weakly interacting 
meson gas and the quark-gluon plasma. 
In the first regime (assuming
for simplicity that we only deal with pions) we obtain for the euclidean
pion propagator at finite inverse temperature $\beta$ ($t>0$): 
\bea 
G_{\pi}^{\beta}(t) \sim {\rm cosh}({\tilde E}_{\pi} ({{\beta}
\over 2} - t )), \\ 
{\tilde E}_{\pi} = E_{\pi} - {{\sum \Delta_n c_n\e^{-\beta E_n}}
\over {\sum  c_n\e^{-\beta E_n}}}
\label{e.pgas3}
\eea
\no with  $E_n$ the energy of the $n-$pion state and 
 $\Delta_n = E_n + E_{\pi} -E_{n+1}$
 measuring the interaction. As long as this interaction is weak we can 
simulate the changes by a shift and possibly a widening of the
peak in the spectral function. 
If the changes are large -- as they should be if the 
quark-gluon plasma regime becomes dominant -- the above 
description breaks down.

These genuine temperature effects should now be distinguished from the 
admixture of excited states in the mesonic channel under consideration.
Our strategy is the following: we try to fix at zero temperature a mesonic source which gives a high projection onto the
ground state. Then we use this source to determine the changes induced by the
temperature on the ground state so defined. From the above
considerations this is a reasonable procedure in the frame of a weakly
interacting meson gas and it is justified as long as the observed changes 
are not too large. The appearance of large changes will then signal the
breakdown of the weakly interacting gas picture and there
we shall  try to compare the observations
with those given by the quark-gluon plasma picture. We stress 
that in this case we do no longer have a good justification to use
 that source as representative of the pion. Our conjecture is that
it still projects onto the dominant low energy structure in the
spectral function but further analysis are necessary to test this
conjecture. In all the figures only statistical errors will be 
displayed since we have no estimation for such systematic effects.

The correlators  investigated are  of the form:
\begin{eqnarray}
 F({\bf P},x,t) =  \sum_{\bf z}\hbox{e}^{i{\bf Pz}}
\sum_{{\bf y_1},{\bf y_2}} w({\bf y_1},{\bf y_2})  
  \nonumber \\
 \langle Tr \left[\gamma_5 S({\bf y_1},0; {\bf z}, t)
\gamma_5 S({\bf y_2},0; {\bf z}+ {\bf x}, t)\right]\rangle 
\label{e.corr}
\eea
\no For practical reasons we do not construct a genuine mesonic source but use 
smeared quark sources with point and exponential ans\"atze:
\be
 w({\bf y_1},{\bf y_2}) = w_1({\bf y_1})w_2({\bf y_2})
\ee
\[
 w({\bf y}) \sim \delta({\bf y}) \ {\rm (point)},\ \ 
 w({\bf y}) \sim {\rm exp}(-ay^p)\ {\rm (exp.)}\nonumber
\label{e.sour}
\]
\no For the exponential source we fix the parameters $a,p$ from the 
wave function (the dependence on ${\bf x}$) measured with point
source at $T=0$. The results of a variational analysis using 
point-point, point-exponential and exponential-exponential 
quark sources indicate that the latter ansatz projects practically
entirely on the ground state. This is well seen from the 
effective mass plots given in Fig. \ref{f.hem72}. Therefore we use for 
our investigations throughout the exponential-exponential source.

\begin{figure}[tb]
\vspace*{-0.28cm}
\leavevmode\psfig{file=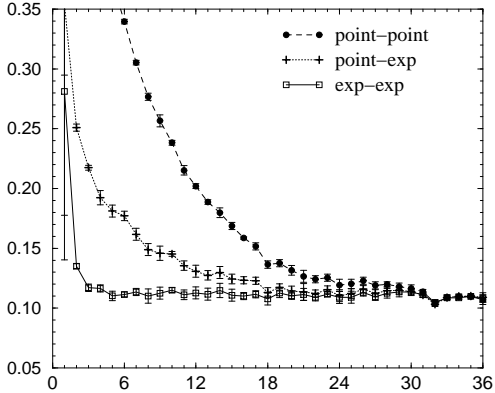,width=\figwidth}
%\leavevmode\psfig{file=f72s.ps,width=1.0\figwidth,angle=-90}
\vspace{-2.6cm} \\
\caption{
Effective pion mass vs $t$ for various sources at $T=0$: point-point, 
point-exponential and exponential-exponential.}
\label{f.hem72}
\vspace{-0.6cm}
\end{figure}

\begin{figure}[tb]
\vspace*{-0.6cm}
%\hspace*{-0.6cm}
\leavevmode\psfig{file=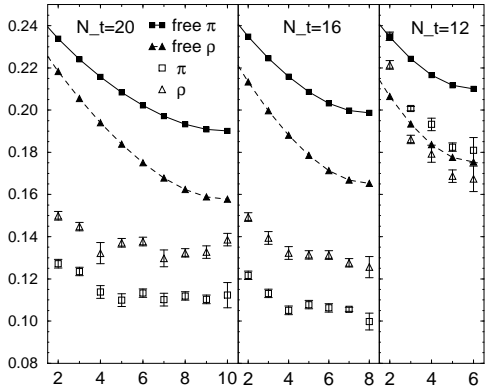,width=1.0\figwidth}
\vspace{-2.3cm} \\
\caption{
Effective $\pi$ and $\rho$ mass at $T\simeq 0.93$, $1.15$ and $1.5T_c$ (open
symbols) vs $t$.
Also shown are the effective masses from the same correlators 
calculated
using free quarks.}
\label{f.hemT}
\vspace{-0.5cm}
\end{figure}

\section{Temperature dependence of the time correlators}

We first present results for the effective mass of the {\it Set-A} data
obtained from the 
$\pi$ and $\rho$ time-propagators at $T \simeq 0.93 T_c$,
$1.15T_c$ and $1.5 T_c$ -- see Fig. \ref{f.hemT}. Comparing with the results at $T = 0$ 
(Fig. \ref{f.hem72}) we notice practically no change at  $T \simeq 0.93 T_c$
and relatively little change at $1.15T_c$. In contradistinction, 
the $1.5 T_c$ plot shows rather large effects: not only does the
effective mass depend strongly on $t$ and becomes significantly larger, but 
the $\pi$ and $\rho$ reverse their positions.

Of course, one should only 
compare   distances up to 6, or about 1.4 fm, the largest distance 
available at  $1.5 T_c$. Performing a fit to the propagators in this
region and using a 
resonance ansatz shows a width developing above $T_c$, leading to
a very flat  structure at $1.5 T_c$.\footnote{It is well known that
due to the KMS condition poles cannot be moved off the real axis 
in an arbitrary way --
see \cite{tgen}, see also \cite{hns1}. It is  possible, however,
to write down an ansatz for a state with non-vanishing width which does not
violate the KMS condition \cite{ks}.
}

Since at $1.5 T_c$ there is every indication that the weakly interacting pion
gas picture is lost we tried to see to what extent the situation is now
compatible with the quark-gluon picture. For this we compare the mesonic
correlators with correlators obtained by using, in the same channels 
and with the same source,
free quark propagators: $S_0$ instead of $S$ in (\ref{e.corr}) --
see eq. (5) in \cite{conf95} (notice that this equation should be corrected at $t=0$
by adding a term $({\rm tanh}(\eta N/2) +\gamma_4)/2A$).  The results are also shown in  Fig.  \ref{f.hemT}. 
The comparison cannot be made quantitative enough, since 
we cannot exactly determine what quark mass we should use for the free quarks.
The data in the figure correspond to massless quarks, higher masses
do not change the general aspect. The conjecture
suggested by the comparison is that
at $1.5 T_c$ a strong quark-gluon component has developed and
the  mesons are in equilibrium with their decay products, quarks and gluons. 

\begin{figure}[tb]
\vspace*{-0.28cm}
\leavevmode\psfig{file=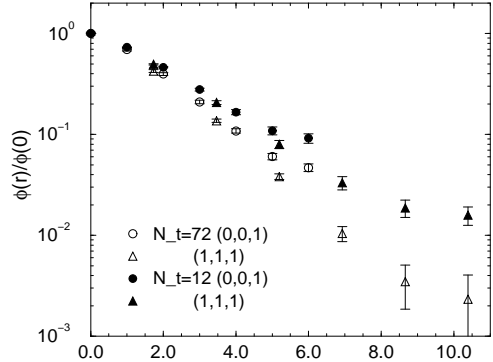,width=\figwidth}
\vspace{-2.8cm} \\
\caption{
The pion ``wave functions" vs $r$ for $\kappa_{\sigma}=0.081$
at $N_t=72$ and $N_t=12$.
They are measured both in (0,0,1) and in (1,1,1) directions. }
\label{fig:wave_func}
\vspace{-0.3cm}
\end{figure}

\section{Screening masses and quark mass dependence}

Further analysis of
the pole and screening masses, and their
dependence on the quark mass are carried
out on the {\it Set-B}, on which three values
of quark mass and higher statistics (30 configurations) are available.

The pole masses $m\equiv m^{(\tau)}$ are extracted from the correlators in 
time (temperature) direction using again smeared quark propagators 
with the exp. - exp. source tuned at $T=0$, as
 described in section 3.
We stress again that at high temperature there is a certain
arbitrariness in considering the state on which this source
projects as the dominant low energy state.
The effective masses   
show a similar behavior with those from {\it Set-A}. For further considerations
we fit the correlators to  single hyperbolic functions 
in the regions: 27-45 ($T=0$), 7-13 ($T\simeq 0.93T_c$), 
6-10 ($T\simeq 1.15T_c$) and 4-8 ($T\simeq 1.5T_c$), which are chosen observing
the effective mass plots.
The results of $\kappa_{\sigma}=0.081$, our heaviest quark mass,
are shown in Fig. \ref{fig:wave_func} and Fig.~\ref{fig:t-dep}.
In Fig. \ref{fig:wave_func} we show the dependence
of the correlators on the distance between quark and anti-quark
at the sink at $T=0$ and at 
$1.5T_c$. 
The wave function is seen to become more flat at  higher 
temperature.

The screening masses $m^{(\sigma)}$ are extracted from the correlators
in z-direction.
At $T=0$, $m^{(\sigma)}=m^{(\tau)}$
is ensured by tuning $\gamma_F$.
At finite temperature, in general the screening mass and the pole mass
do not coincide.
Since the physical distance in the space direction is 
reasonably large we extract the screening masses from  
correlators with {\it point-point} source
in the range 5-7 at all temperatures (although at $T=0$ 
no clear plateau is yet observed, the result agrees within a few
percents with the mass obtained with a Wuppertal source).
At  $T\simeq 0.93T_c$ the z-correlators are very similar to
those at $T=0$.
Above $T_c$, the effective masses seem to reach the plateau beyond $z=4$.
The results for $m^{(\sigma)}$ at $\kappa_{\sigma}=0.081$ are also shown
 in Fig.~\ref{fig:t-dep}.

\begin{figure}[tb]
\vspace*{-0.08cm}
\leavevmode\psfig{file=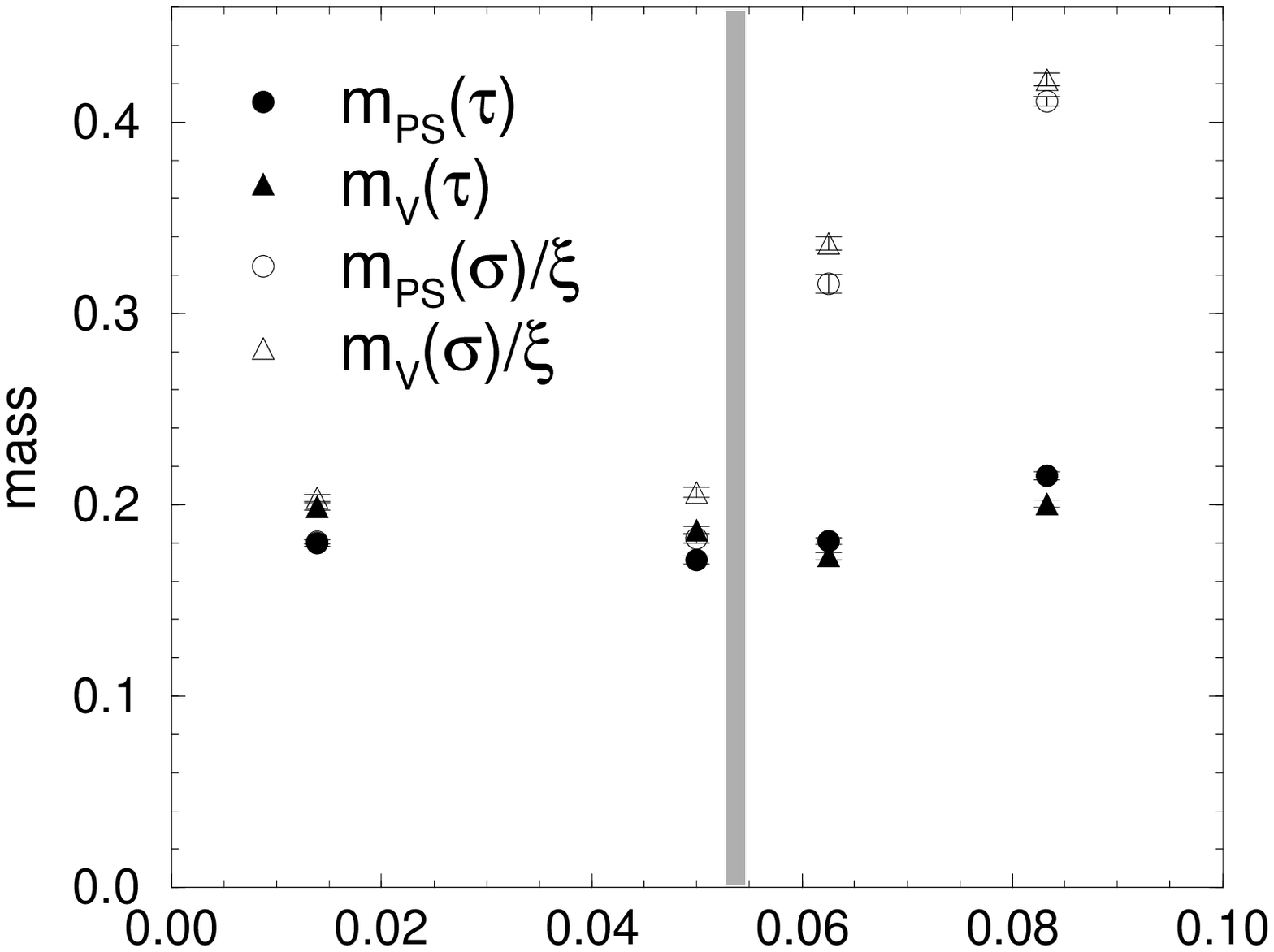,width=\figwidth,height=\figheight}
\vspace{-1.3cm} \\
\leavevmode\psfig{file=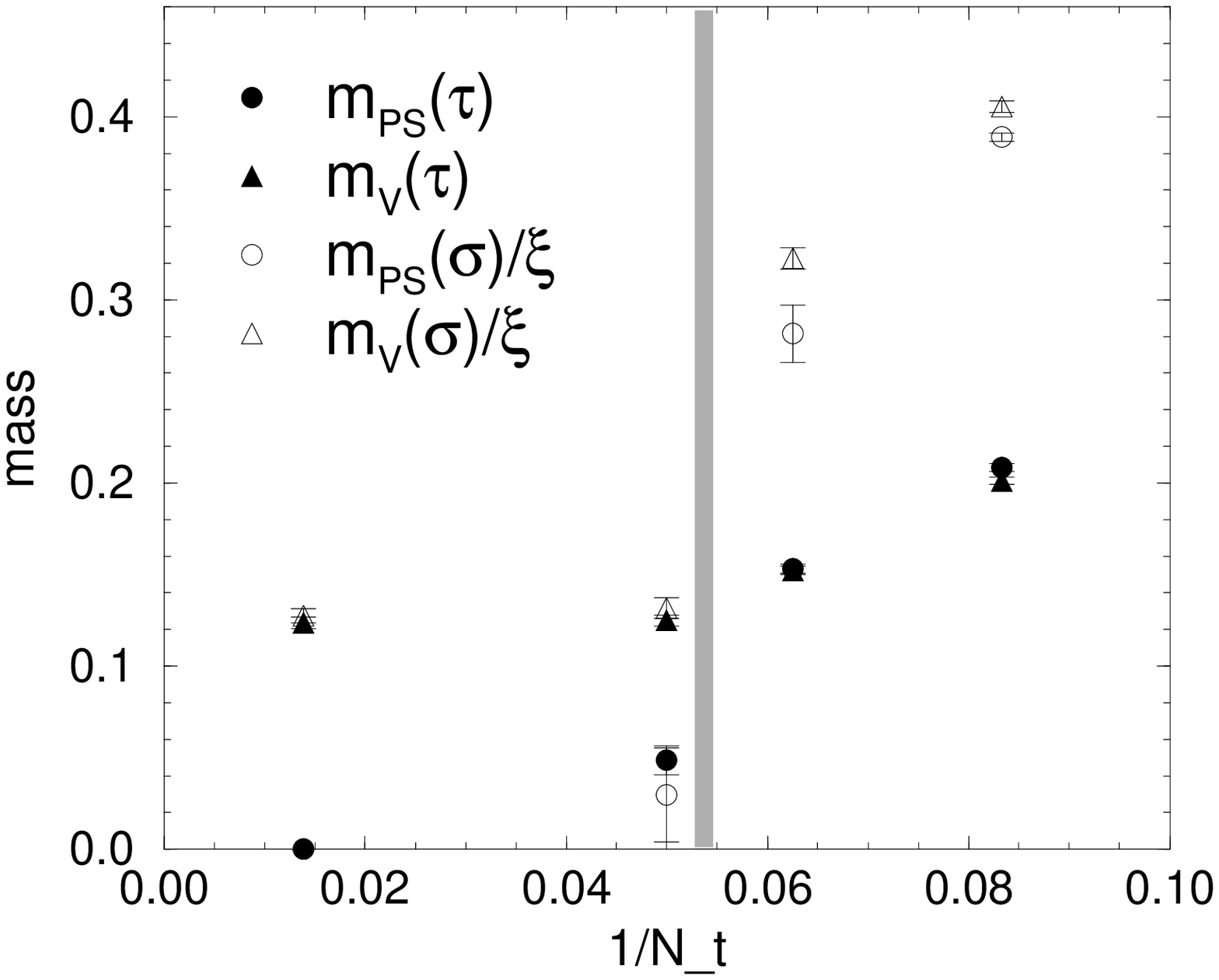,width=\figwidth,height=\figheight}
\vspace{-2.4cm} \\
\caption{
Temperature dependence of the pole and the screening masses for
$\kappa_{\sigma}=0.081 $(top figure), and at the chiral limit (bottom).
The vertical gray lines roughly represent the critical temperature.  }
\label{fig:t-dep}
\vspace{-0.5cm}
\end{figure}

Both pole and screening masses are extrapolated to the 
chiral limit from the 3 quark masses analyzed. The extrapolation is
done using  $\kappa$ of eq. (\ref{e.kappa}).
Below the critical temperature
the pion masses squared $m_{ps}^2$ are extrapolated,
in all other cases, the masses themselves 
are extrapolated linearly in $1/\kappa$.
The extrapolation of the pole masses at $T=0$ gives
the critical value $\kappa_c=0.17138(20)$ and 
$a_{\sigma}^{-1}= 1.18 $ GeV from the $\rho$ mass using $\xi=5.3$.
The result of the screening mass extrapolation is almost the same
-- see Fig~\ref{fig:chext}.
The large difference between $a_{\sigma}^{-1}(\rho)$ and $a_{\sigma}^{-1}$
from the heavy quark potential (see the table)
indicates significant discretization
effects for our lattices as expected.
The results are summarized in the second plot in Fig.~\ref{fig:t-dep}.
We find large differences between the pole and the screening masses
above $T_c$ both at finite quark masses and in the chiral limit.
We should like to point out that a similar behavior is
 obtained in an effective model approach with NJL Lagrangian
\cite{HK94}.

\begin{figure}[tb]
\vspace*{-0.28cm}
\leavevmode\psfig{file=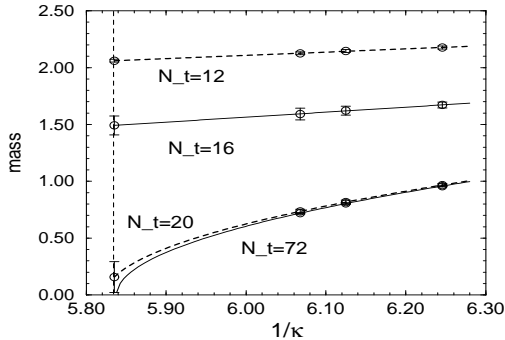,width=\figwidth,height=1.18\figheight}
\vspace{-2.4cm} \\
\caption{
Chiral extrapolation of the pion screening masses at various $T$.}
\label{fig:chext}
\vspace{-0.5cm}
\end{figure}

\section{Topological properties}

To determine the topological properties of the configurations we use 
improved cooling -- see \cite{mnp}. Unlike usual cooling with Wilson
action this method preserves instantons at scales larger than a certain
dislocation threshold $\rho_0$. 
For  anisotropic lattices the improved action is 
modified accordingly ensuring isotropic distributions for
the sizes and distances of instantons at $T=0$.
 From earlier investigations we expect
a scale of about 0.4 - 0.7 fm to be relevant for the topology of $SU(3)$. 
With a dislocation threshold  $\rho_0 \sim 2a \simeq 0.5$ fm the 
lattices used here are too coarse for a detailed analysis. 
Nevertheless we obtain reasonable data for the
susceptibility in agreement with Witten-Veneziano formula below $T_c$
and dropping to zero at high temperature, in
agreement with other analysis on finer lattices. An
interesting quantity is the ratio $R(k) = Q_t^2(kV)/Q_s^2(kV)$ 
between the charge 
in subvolumes $kV$ obtained by slicing the 
lattice in either a spatial or a temporal direction. The temperature effect
shows up in an anisotropy between temporal and spatial slicing which sets in
when the dominant sizes begin to feel the limited temporal extension.
Just below $T_c$ $R(1/2)$ in the charge zero sector is about 0.7, while 
above $T_c$ (where all configurations have $Q=0$) 
$R(1/2)$ grows fast to about 3. This behavior is compatible with
a random distribution of instantons of rather large sizes below $T_c$
and a tendency to form pairs above $T_c$. Generally,
there seems to be a qualitative difference between the instanton 
population below and above $T_c$: at similar densities of pairs the 
instantons of the low temperature are reasonably 
well fitted by the 't Hooft ansatz, while at high temperature we seem to deal with topological charge 
fluctuations which hardly can accommodate the continuum picture of
instantons. A more detailed account of these features will be given elsewhere.

\section{Conclusions and outlook}

The present analysis has produced results
about the temperature effects on the vacuum and hadronic
properties in quenched QCD which are consistent with the following picture:
below $T_c$ the changes are small and gradual while above $T_c$ the changes 
increase strongly (but not abruptly) with the temperature; at temperatures 
of about $1.5 T_c$ the mesons have become  unstable and are strongly
interacting
with a significant quark-gluon plasma component; the changes in the
vacuum structure follow a similar pattern and there is indication of 
close range opposite charge correlation (although there are no dynamical fermions).

This picture needs to be further tested in order to remove
the uncertainties still affecting the present analysis. This concerns
particularly the
question of an unambiguous definition of the hadron states at high temperature.
In the further developments we shall also try to extract directly 
information about the
spectral function of the hadronic correlators \cite{hsp}, use larger lattices such that 
we can both have larger physical distances and
a smaller lattice spacing (in spatial directions) for more 
refined studies of the topology, and aim, of course, to an improved statistics.     

The calculations have been done on AP1000 at Fujitsu Parallel 
Computing Research Facilities and Paragon at INSAM, Hiroshima Univ.


\begin{thebibliography}{9}
 
\bibitem{tgen} N.P.~Landsman and Ch.G.~van~Weert, Phys. Rep. {\bf 145} 
(1987) 141.

\bibitem{biel1} QCD-TARO: M.~Fujisaki et al.,
in {\it Multi-scale Phenomena and their
Simulation}, F.~Karsch, B.~Monien and H.~Satz eds., World Scientific 
(Singapore 1997) 208. 

\bibitem{Sakai98}
 S.~Sakai, A.~Nakamura and T.~Saito, these proceedings.

\bibitem{TARO96}
 QCD-TARO: M.~Fujisaki et al.,
  Nucl. Phys. B (Proc.Suppl.) {\bf 53} (1997) 426.

\bibitem{hns1} T.~Hashimoto, A.~Nakamura and I.-O.~Stamatescu,
 Nucl. Phys. {\bf B400} (1993) 267.

\bibitem{ks} J.~Kupsch and I.-O.~Stamatescu, work in progress.

\bibitem{conf95} QCD-TARO: M.~Fujisaki et al.,
in {\it Confinement 95}, H.~Toki et al eds., World Scientific 
(Singapore 1995) 129.

\bibitem{HK94}
 T.~Hatsuda and T.~Kunihiro,
  Phys. Rep. {\bf 247} (1994) 221.

\bibitem{mnp} Ph.~de Forcrand, M.~Garc{\'\i}a P\'erez and I.-O.~Stamatescu,
Nucl. Phys. {\bf B499} (1997) 409.

\bibitem{hsp}  QCD-TARO: Ph.~de~Forcrand et al.,
Nucl. Phys. B (Proc.Suppl.) {\bf 63} (1998) 460.


\end{thebibliography}
\end{document}